\def\ref{\noindent\hangindent=3pc\hangafter=1}
\def\etal{et\ al.\ }
\def\apj{ApJ,~}
\def\mnr{MNRAS,~}
\def\apjs{ApJS,~}

\def\Gi{{\it Ginga}}

\def\cm{{\rm\thinspace cm}}
\def\erg{{\rm\thinspace erg}}

\def\s{{\rm\thinspace s}}

\def\ergpcmsqps{\hbox{$\erg\cm^{-2}\s^{-1}\,$}}

\def\ergps{\hbox{$\erg\s^{-1}\,$}}

\def\psqcm{\hbox{$\cm^{-2}\,$}}

\def\ref{\par \noindent \hangindent=3pc \hangafter=1}

\def\eg{{ e.g.\ }}

\def\spose#1{\hbox to 0pt{#1\hss}}
\def\approxlt{\mathrel{\spose{\lower 3pt\hbox{$\sim$}}
	\raise 2.0pt\hbox{$<$}}}
\def\approxgt{\mathrel{\spose{\lower 3pt\hbox{$\sim$}}
	\raise 2.0pt\hbox{$>$}}}
\mathchardef\twiddle="2218

\def\multleft#1{\hbox to size{\vbox {\halign {\lft{##}\cr #1}}\hfill}\par}
\def\multright#1{\hbox to size{\vbox {\halign {\rt{##}\cr #1}}\hfill}\par}
\def\<{\thinspace}

\documentstyle[11pt]{article}
\begin{document}
\begin{center}
{\LARGE Fluctuations in the diffuse X-ray background observed 
with {\it Ginga}}
\end{center}
\vspace{1.0cm}
\begin{center}
{\Large J.A. Butcher$^1$, G.C. Stewart$^1$, R.S. Warwick$^1$,
 A.C. Fabian$^2$, F.J. Carrera$^3$, X. Barcons$^3$, 
 K. Hayashida $^{4,5}$,  H. Inoue $^4$, T. Kii $^4$ }
\end{center}

{ 1. Department of Physics and Astronomy, Leicester University, 
University Road, Leicester, LE1 7RH, UK. } \newline
{ 2. Institute of Astronomy, Madingley Road, Cambridge CB3 0HA, UK.} \newline
{ 3. Instituto de F\'\i sica de Cantabria (Consejo Superior de Investigaciones 
Cient\'\i ficas-- Universidad de Cantabria), 39005 Santander, Spain.} \newline
{ 4.  Institute of Space and Astronautical Science, 3-1-1 
Yoshinodai, Sagamihara, Kanagawa 229, Japan.} \newline
{ 5.  Department of Earth and Space Science, Osaka University, 
Machikaneyama, Toyonaka, Osaka 560, Japan.} \newline

{\Large {\bf Abstract}} \newline
We present {\it Ginga} measurements of the spatial fluctuations in the
diffuse X-ray background. When combined with earlier results, the new data
constrain the extragalactic $\log N - \log S$ relation in the 2--10~keV 
energy band to a form close to the Euclidean prediction over the flux range
$10^{-10} - 5\times10^{-13} \rm~erg~cm^{-2}~s^{-1}$. The normalisation of 
the 2--10 keV source counts is 
a factor $2-3$ above that derived in the softer 0.3--3.5 keV band from the 
{\it Einstein} Extended Medium Sensitivity Survey if a spectral conversion is 
assumed which ignores X-ray absorption intrinsic to the sources. Both this 
result and
the spectral characteristics of the spatial fluctuations are consistent with
relatively low-luminosity active galaxies ({ i.e.} $L_{X} < 10^{44}~ 
\rm erg~s^{-1}$) dominating the 2--10 keV source counts at 
intermediate flux levels. We also use the `excess variance' of the 
fluctuations to constrain possible clustering of the underlying discrete 
sources.

{\begin{center}
\end{center}

\newpage
 
\section{INTRODUCTION}

Since the discovery of the diffuse X-ray background (XRB) there has been 
considerable debate as to whether its origin lies solely in the
integrated emission of known types of extragalactic X-ray source, such as 
Seyfert galaxies and QSOs, or whether a significant contribution might come
from a hitherto undiscovered population of discrete X-ray sources.
The alternative, that much of the XRB might originate in 
diffuse emission from a putative intergalactic medium, 
has been effectively rejected through measurements of the microwave
background spectrum using the $COBE$ satellite (Mather \etal 1990; Barcons,
Fabian \& Rees 1991).

In modelling the XRB
in terms of discrete sources a continual frustration is the fact
that the surface brightness and spectral form of the XRB are well determined 
only above 3~keV, whereas the most detailed studies of the source 
populations have been carried out in the soft X-ray 
band (through the exploitation of the imaging systems on {\it Einstein},
{\it EXOSAT} and {\it ROSAT}). In the 2--10 keV band, which contains a
significant fraction of the XRB energy density, our knowledge of 
the X-ray luminosity and spectra of individual sources is restricted mainly to 
the relatively bright sources which appear in the all-sky catalogues of 
{\it Uhuru}, {\it Ariel 5} and {\it HEAO-1}. However,
the statistical properties of sources at much fainter fluxes are accessible
through studies of spatial fluctuations in the XRB.

Radio astronomers were the first to recognise that, by studying the
beam-to-beam variations in the sky background in directions away from bright
catalogued sources, it was possible to derive information on sources too faint
to be resolved  individually. The theoretical basis for such fluctuation
studies is well documented (Scheuer 1974; Condon 1974).  Provided source
clustering  effects can be neglected (see Barcons 1992), the form of the
background  fluctuations distribution, the so-called $P(D)$ distribution, is
governed  solely by the $\log N - \log S$  relation of the discrete sources
which produce the fluctuations and by the  normalised beam response of the
detector. It can be shown that,  for simple power-law forms of $N(S)$, the
shape of the $P(D)$ distribution  depends on $\gamma$, whereas the width of the
$P(D)$ distribution scales as  $(K\Omega_{e})^{1/\gamma-1}$, where $\Omega_{e}$
is the effective  beamsize (see Condon 1974). 
We use the differential form of the 
$\log N - \log S$ relation, in which the number of 
sources per steradian, in the flux range from $S$ to $S+dS$, is given 
by:
 
\[
N(S)dS = KS^{-\gamma}dS
\]

Here $K$ is the normalisation of the differential source counts 
and $\gamma$ the power-law slope; note that $\gamma=2.5$ corresponds to the 
Euclidean form of the source counts.

The standard analysis
approach is one  of model-fitting,  in which a predicted $P(D)$
curve is computed and compared to the observed distribution. The properties of
the underlying $\log N - \log S$ relation  can then be inferred through the
process of minimising the differences  between the predicted and observed
fluctuation distributions. Although the  spatial fluctuations are due to the
combined influence of sources over a  very wide flux range, in practice most
information is extracted at flux  levels corresponding to a surface density of
about 1 source per beam (Scheuer  1974). It follows that 
the narrower the detector beam the fainter the flux level reached. An important
proviso is, of course, that the fluctuations must be measured with adequate
signal-to-noise ratio  for useful information to be obtained.
 
In the present paper we show that measurements of spatial fluctuations 
by {\it Ginga} provide constraints on the 2--10 keV source counts and source
clustering at flux levels corresponding
to a surface density of extragalactic X-ray sources of $\sim 1$ object per 
square degree. Similar levels are encompassed in the {\it Einstein} 
Observatory Extended 
Medium Sensitivity Survey (EMSS) (Gioia \etal 1990b) and a comparison is 
made between the source 
count estimates in the medium energy and soft X-ray regimes ({ i.e.} the
2--10~keV and 0.3--3.5~keV bands). We also 
comment on the spectral characteristics of the spatial fluctuations measured 
by {\it Ginga}.  

There have been several previous studies of XRB fluctuations as measured
in the medium energy X-ray band (e.g. Fabian 1975; Schwartz, 
Murray \& Gursky 1976; Pye \& Warwick 1979; Shafer 1983). Of these, 
the most definitive result is from the {\it HEAO-1} A2 survey in which 
the XRB data were derived from detectors with 
collimated fields-of-view of $3^{\circ}\times3^{\circ}$ and 
$3^{\circ}\times1.5^{\circ}$ (Shafer 1983). Here we present
measurements of spatial fluctuations in the XRB made with the 
Large Area Counter (LAC) on {\it Ginga}. Since the collimated beam of the LAC 
is only $1^{\circ}\times2^{\circ}$, the {\it Ginga} measurements are
potentially more sensitive than those from {\it HEAO-1} A2.
Full details of the LAC, which had a total effective area 
of $\sim$ 4000 cm$^2$, several times larger than that of the
{\it HEAO-1} A2 experiment, are given in Turner \etal (1989).
Preliminary results from this study have been reported by Warwick \& Stewart
(1989), Hayashida (1990) and Warwick \& Butcher (1992).  {\it Ginga} 
observations (the same data as used here) have also been used to 
study the angular autocorrelation function 
of the X-ray background (Carrera et al. 1991).

The paper is organised as follows.  Section 2 presents the estimates of source
counts based on the analysis of the {\it Ginga} fluctuations.  Upper limits to
the excess variance required in the fitting procedure are used to constrain
source clustering.   The spectrum of the fluctuations is also analyzed. 
Section 3 is devoted to the discussion of these results and
in particular to the comparison of the inferred source counts with the EMSS.
Section 4 summarizes our results. 

\section{THE {\it GINGA} MEASUREMENTS}

\subsection{The observations}

The data used in the fluctuation analysis come from two types of observation:
dedicated raster scans of the background and observations taken as part of 
the normal background monitoring programme; in total the dataset 
comprises 132 independent pointings. 
The data were all taken in MPC1 mode (Turner et al. 1989) and we use only
data from the top layer of the LAC (separation of data from the top and mid 
layers of the LAC gives an improved  signal-to-noise ratio for weak sources).
In all of the observations the angle between the LAC pointing direction 
and the sun was greater than $90^{\circ}$, thus excluding any possibility
of contamination of the data by solar X-rays.

Since \Gi \ had no facility for simultaneous background measurement during 
a source observation, a typical \Gi \ observation consisted of one day 
pointing at the source followed by a further day pointing at a nearby region 
of `blank' sky. Over the lifetime of the mission there thus accumulated a
large database of potentially useful background observations. Here we
utilize 70 such observations of independent sky regions made between 1987
June and 1989 August. We selected only observations with $|b| >
25^{\circ} $. Where there existed multiple pointings in the same
direction we chose the one with the largest exposure. The typical
exposure of an observation is $\sim 20000$ s. Figure 1 shows the
distribution of these observations over the sky. It should be noted that
the distribution of these pointings is neither uniform nor random but
follows, to first order, the distribution of X-ray sources above the \Gi
\ limiting sensitivity.

The second source of data is from three sets of dedicated background 
observations performed in 1988 January, 1988 July and  1989 March. In each
case a set of independent regions of sky were observed at the 
rate of one region per satellite orbit giving a typical exposure 
time per pointing of $\sim 1500$ s. The three areas of sky observed in this
way were situated near the north and south galactic poles and the north 
ecliptic pole.

\subsection{Extraction of the background fluctuations}

The LAC registered counts from charged particles and other events as well as
from cosmic X-rays. Both the count rate and spectrum of the LAC internal
background varied with time in a complex manner (see Hayashida et al.
1989).  Over a period of $\sim 24$ hr most of the variation of the 
background could be explained by the change in cutoff rigidity with 
satellite altitude and by
radioactive decays induced by the passage of the satellite close to the
South Atlantic Anomaly (SAA). In addition a 37-day periodicity was present 
as a result of the precession of the satellite orbit which caused the height 
of the daily passage through the SAA to vary. There was also a very long 
term decrease in the
background which was related to the gradual decay of the satellite orbit. The
internal background of the LAC was modelled using a development of the methods
described in Hayashida et al.(1989), although the exact method employed
was slightly different for the two different types of data.

Each set of dedicated observations lasted only 1--2 days, so the long term
variations could be ignored. For each dataset we fitted a time series 
model of the form:

\[
I(E,t) = a_1(E) + \sum_{i=2}^{6} a_i(E) \times x_i(t)
\]

where $I(E,t)$ is the LAC count rate as a function of energy channel 
and time, $x_i(t)$ are five time-dependent parameters derived from
LAC housekeeping data and $a_i(E)$ are scaling coefficents which
represent the spectral forms of the corresponding background components. 
More specifically two of the $a_i$ coefficients represent contributions which 
scale with the SUD (counts Surplus to Upper Discriminator) and with the cutoff 
rigidity, whereas the other three represent the spectra of
induced radioactive decays with empirically-derived 
e-folding times of 12.5, 1 and 0.5 hours. The time-independent 
term $a_1(E)$ is the average over the dataset of the cosmic diffuse 
background plus any residual particle background in the detector.
The best fit values of $a_1$ and the five $a_i$ coefficients for each
set of dedicated observations were used to create a background model 
for that dataset. The deflection or fluctuation level was then
determined for each individual pointing direction as the difference between 
the observed count rate and the modelled value.

The equivalent procedure for the observations drawn from the \Gi \
database was slightly more complicated. As the observations were 
spread over a longer time period it was necessary to model the variations 
in internal background caused by the 37-day
precession of the satellite orbit (Hayashida et al. 1989). Ideally such 
modelling requires many observations spread over all phases of the 37-day
period. There was also the problem of the long term drift of the internal 
background induced by the decay of the satellite orbit.
The latter proved difficult to incorporate directly in the fitting procedure
and, as a first approach to minimising any long term drift 
whilst attempting to optimise the modelling of the 37 day component, 
the 70 observations were simply divided into two datasets containing 36 
and 34 pointings, each of which spanned roughly 13 months. 
A background model of the form discussed earlier was then applied to
each of these datasets except that the $a_i$ coefficients modelling the 
induced radioactivity incorporated a (best-fitting) 37-day variation. 
The fluctuation level for a particular observation was again determined 
as the difference between the observed count rate and that derived from 
the background model for the dataset. Finally a check  was made for
evidence of any obvious trends in the time-ordered fluctuation 
measurements from each dataset; however, no significant long-term effects 
were apparent, thus validating the analysis procedure.

\subsection{Analysis of the $P(D)$ distributions}

The $P(D)$ distributions measured by {\it Ginga} in two energy
ranges (2--4~keV and 4--12~keV) are shown in Fig. 2.
The rms widths of the distributions correspond to approximately
$\sim 0.4 \, \rm ct \, s^{-1} $ which corresponds to roughly 5  per cent
 of the predicted count-rate from the XRB in the two bands 
(using predictions based on the XRB spectrum reported by Marshall \etal 1980).

We fit the observed $P(D)$ curves using trial number
count relationships of power-law form using standard Fourier 
transform techniques based on the method of Condon (1974). 
Full account is taken of the beam profile 
of the LAC collimators. The $\log{N} - \log{S}$ relationship is
truncated when the integrated intensity exceeds the total
background strength, although as explained earlier, the shape of the $P(D)$ is
not sensitive to fluxes where the $N(S)$ curve predicts many sources per beam. 
The effects of counting statistics and residual errors in the fitting procedure
described above
are included by convolving the model $P(D)$ curve with a Gaussian of standard 
deviation $\sigma\approx 0.1\,
{\rm ct}\, {\rm s}^{-1}$. This is the appropriate statistical error for
the short $1-2 \times 10^3$ second scan observations while for the longer
pointings the value is dominated by the systematic errors
in the background modelling procedure. For the database pointings the
value of $\sigma$ was estimated by comparing repeated observations 
of the same region of sky. The results are not sensitive to the exact value of
$\sigma$ within appropriate limits. (We note that this is the first time that
XRB fluctuation measurements have been made where the P(D) noise is
several times the statistical error on individual samples.)

\subsection{Constraints on the $\log N - \log S$ relation}

The observed fluctuation distributions were fitted with model $P(D)$ 
curves, via a process of $\chi^2$ minimisation. (The results thus obtained 
were entirely consistent with those produced by an alternative approach, 
namely the maximum likelihood method - see the next section). This fitting 
procedure lead to the 
constraints on the power-law slope, $\gamma$, and normalisation, $K$,
of the $\log N - \log S$ relation shown in Fig. 3. The parameters of 
the best-fitting models, for which the predicted P(D) curves are shown 
in Figure 2,  are listed in Table 1, where the quoted errors correspond
to 90 per cent confidence for one interesting parameter. Table 1 also 
gives the best-fitting normalisations for the $\gamma = 2.5$ case, 
that is a Euclidean slope for the source counts. Finally Table 1 also 
gives the normalisations converted to 2--10 keV flux units using 
a conversion factor based on the response of {\it Ginga} to a power-law 
spectrum with an energy index of $\alpha = 0.8$ (we show later that this 
is an appropriate value for $\alpha$). 

The best-fit slope for both energy ranges are consistent with each other
and with the Euclidean value of $\gamma = 2.5$.
The inferred Euclidean 2--10 keV source count normalisations in the 
two energy bands are also in excellent agreement with
each other and with the normalisation of the resolved-source counts 
reported by Piccinotti et al. (1982).

\subsection{Excess variance}

So far we have explicitly neglected source clustering in the computation
of the model $P(D)$ distribution.  It is customary (Fabian \& Rees 1978;
Shafer 1983) to mimic the effects of source clustering by convolving the
predicted distribution of fluctuations for unclustered sources with a
Gaussian of standard deviation $\sigma_{\rm excess}$ and then to try to
measure or constrain this parameter via a comparison with the
observational data.

In order to study the allowed range for the excess variance, we have used 
a maximum likelihood method, since checks revealed that, in this case,
binning the data and minimizing the $\chi^2$ function did appear to give
systematically higher values (by up to 20 per cent) of $\sigma_{\rm excess}$ 
than with the maximum likelihood approach.
We have computed the likelihood function in $\left(K,\gamma,\sigma_{\rm
excess}\right)$ parameter space.
  We then consider $K$ and $\gamma$ as non-interesting parameters and for
all their values we select the maximum likelihood  as a function of
$\sigma_{\rm excess}$ only.  This function is then normalized to give a 
measure of probability.

We find that the value of $\sigma_{\rm excess}$ that maximizes the likelihood 
function is less than $0.05$~ct~s$^{-1}$, which is of the same order as the 
errors in the fluctuation measurements. Thus there is no clear  evidence for 
effects arising from source clustering. At the one sigma level 
$\sigma_{\rm excess}<0.17$~ct~s$^{-1}$ and at two sigma confidence 
$\sigma_{\rm excess}<0.3$~ct~s$^{-1}$. In relative terms those upper limits
correspond to ${\Delta I\over I}<2$~per cent and ${\Delta I\over I}<3.8$~per 
cent. It must be stressed that the two sigma upper limit corresponds 
to a situation in which there is a very low surface density source 
population  and most of the width of the $P(D)$ curve comes from the 
excess variance. 
These results on source clustering are entirely consistent with the results of
Carrera et al. (1991) who use the same data  reported here to calculate the 
ACF of the X-ray background and obtain only upper limits. 

The two sigma upper limit ${\Delta I\over I}<3.8$~per cent poses some
interesting cosmological constraints. In terms of Poisson fluctuations this
limit implies a minimum of 345 sources deg$^{-2}$, whereas Shafer (1983) 
obtained a limit of 2 per cent for a 25~deg$^2$ beam, thus
requiring a minimum of only 100 sources~deg$^{-2}$.  
The present measurements pose
stringent constraints on any large-scale inhomogeneities of the X-ray Universe 
(see Fabian \& Barcons 1992).
This is emphasized in Fig.~4 where the upper bounds to 
${\delta\rho\over\rho}$ (assuming that X-rays trace mass on large scales)
implied by the lack of significant excess variances are shown. It is assumed
that the sources that dominate the fluctuations in the LAC are concentrated at
a mean redshift of about $z=0.5$ with a spread $\Delta z=0.5$, 
which is very roughly consistent with the EMSS redshift distribution. 
The present data provide the best direct evidence for the linearity of 
the fluctuations in the Universe on scales greater than several tens of Mpc 
at those early redshifts.

\subsection{The spectrum of the fluctuations} 

A further potentially valuable source of information on source properties at
relatively faint flux levels is provided by the spectral characteristics of the
spatial fluctuations in the XRB. We have extracted the spectral form of the
fluctuations by subtracting the average spectrum of ``negative'' fluctuations
from that of  ``positive'' fluctuations. (The spectral parameters are
very mildly dependent on the definition of ``negative'' and ``positive'' 
fluctuations).

The resulting spectrum  was then analysed by a standard model fitting 
procedure. We assume a power-law spectral form and initially fit the data
over the 2--11 keV range. The fit was formally acceptable and the derived
spectral parameters were $\alpha = 0.82^{+0.04}_{-0.03}$ and
$N_H < 0.9 \times 10^{21}\rm~cm^{-2}$ (errors are $ 1\sigma$ for 2
interesting parameters). Since the fluctuations are dominated by sources 
with a surface density of roughly one per beam, this reflects the 
average spectrum of sources producing about 10 per cent of the 2--10 keV XRB. 
It is notable that the fluctuation spectrum has a
power-law slope which is typically of Seyfert 1 galaxies 
in the 2--10 keV band.
Of interest, however, is an extrapolation of the 
best fitting spectrum to higher energies.
This reveals an apparent hardening of the spectrum  above 10 keV.
Indeed, the simple spectral model described above did not yield a
statistically acceptable fit over the 2--18 keV band ($\chi^{2}/ \nu =
39.5/24$). In this case the derived spectral parameters were 
$\alpha = 0.80^{+0.03}_{-0.04}$ and
$N_H < 0.5 \times 10^{21}\rm~cm^{-2}$ but, in view of
the bad fit, clearly these should be treated with caution. 

The apparent ``hard excess''  is reminiscent of the features
apparent in spectra of Seyfert 1 galaxies,  which have been attributed 
to a Compton reflection bump (e.g., Pounds \etal 1990). In fact the
inclusion of a Compton reflection component in the current analysis 
results in a significant improvement in the 2--18 keV spectral fit
($\chi^{2}/ \nu = 26.7/24.3$). In this case the spectral
parameters were $\alpha=0.90^{+0.10}_{-0.07}$, $N_H < 1.5 \times
10^{21}\rm~cm^{-2}$ with the fraction of reprocessed flux being $\sim $
50\%. Figure 5 shows the measured spectrum of the fluctuations
and the best fitting model including a reflection component.

\section{DISCUSSION}

\subsection{The 2--10 keV X-ray source counts} 

The 2--10~keV source counts for bright ($S > 
3 \times 10^{-11}\rm~erg~cm^{-2}~s^{-1}$)
extragalactic sources have been determined from the {\it Ariel 5}, 
{\it Uhuru} and
{\it HEAO-1} A2 all-sky surveys (Warwick \& Pye 1978; Schwartz 1979; 
Piccinotti { et al.} 1982). There is good agreement between the three 
surveys on the normalisation of the source counts, which is hardly surprising 
given the fact that the
small ``statistical samples'' derived from each survey (typically
$\sim 60$ sources in each case) overlap to a considerable extent. 
The derived slope of the bright source counts is marginally 
steeper than the Euclidean value, but this appears to be due to the 
chance distribution of a few nearby sources rather than being of direct 
cosmological 
significance (Piccinotti { et al.} 1982). At fainter levels we must
rely on spatial fluctuation measurements. This is illustrated in 
Fig. 6 which shows the form of the 2--10 keV differential source
counts derived from the {\it Piccinotti} discrete source 
sample and from both the {\it HEAO-1} A2 (Shafer 1983) and present {\it Ginga}
fluctuation measurements.

The {\it HEAO-1} A2 fluctuation measurements best 
define the source counts at flux levels about a factor 6 below the threshold 
of the {\it Piccinotti} sample, whereas the new {\it Ginga} results
provide additional constraints at flux levels {\it a factor of 30 below} the
{\it Piccinotti} limit, a flux which will only be directly probed by 
imaging medium energy telescopes.

The discrete source and spatial fluctuation data taken together
constrain the $\log N - \log S$ relation to a form close to
the Euclidean prediction over almost 3 decades of flux.  
Specifically the 2--10 keV differential counts are consistent with:

\[
N(S)=2\times10^{-15}S^{-2.5}\, ({\rm erg}\, {\rm cm}^{-2}\, {\rm
s}^{-1})^{2.5}\, {\rm sr}^{-1}
\]

over the range $S = 10^{-10} - 5\times 10^{-13}\rm~erg~cm^{-2}s^{-1}$
(2--10 keV).

We note that although the fluctuation measurements include a
contribution from Galactic sources (which are, of course, 
specifically excluded from the 
discrete source sample), the net effect of Galactic sources,   
particularly at the lower flux levels, will be small ($^{<}_{\sim} 10$ per cent
contribution to $K$ at $S=10^{-12}\rm~erg~cm^{-2}~s^{-1}$).
Similarly the effect of possible large-scale structure in the XRB, for 
example due to Galactic diffuse emission and the 
Compton-Getting anisotropy, is unlikely to introduce a
significant bias (Shafer 1983; Carrera { et al.} 1991). As we noted 
earlier, source clustering might contribute to an overestimate of the 
source counts, but it would have to be implausibly strong to seriously 
affect our conclusions (Carrera et al. 1993).

The fact that the extragalactic source counts have a slope close to the 
Euclidean value necessarily implies some cosmological 
evolution of the source population, since the effects of relativistic
geometry will be quite pronounced even for the modest redshifts which 
correspond to source luminosities typically in the range 
$L_{X} = 10^{42-45}\rm~erg~s^{-1}$.   

\subsection{Comparison with soft X-ray source counts} 

The {\it Ginga} fluctuation studies extend the 2--10~keV source counts
to flux levels encompassed in the 0.3--3.5~keV band by the  
EMSS (Gioia { et al.} 1990b) and it is clearly of  interest to 
compare the $\log N - \log S$ relations derived in the two spectral regimes.

In principle such a comparison requires knowledge of the spectral 
properties of the contributing source populations over a wide energy range 
(0.3--10 keV). However, a convenient simplification (e.g. Boldt 1988) is to 
define an average spectral conversion factor $f$:

\[
f  =  {S_{2-10}\over S_{0.3-3.5}} 
\]

where $S_{2-10}$ is the flux measured in the 2--10 keV band 
and $S_{0.3-3.5}$ is the corresponding value in the 0.3--3.5 keV band. 

If a particular value of $f$ applies to the whole source population then
the differential source count normalisation in the two bands are related 
by:

\[
K_{2-10}  = f^{\gamma-1} \times K_{0.3-3.5}
\]

For a source with a power-law spectrum of
spectral energy index $\alpha=0.7$ (the so-called canonical spectral slope 
for AGN) and  negligible low-energy absorption (we actually assume a minimum
line-of-sight column density of $N_H = 3 \times 10^{20}\rm~cm^{-2}$), 
then $f\sim 1$. Figure 7 shows the effect on $f$ of varying the 
power-law slope and intrinsic absorption.
Using $f=1$, the extragalactic $\log N - \log S$ relation 
from the EMSS (Gioia { et al.} 1990b) transforms to the dotted
line in Fig. 6. Clearly the above assumptions lead to the conclusion that our
fluctuation analysis produces a significant 
({ i.e.}, a factor $2-3$) overestimation of the soft X-ray source counts, as
measured by the {\it Einstein} Observatory.

To reconcile the EMSS and 2--10 keV source counts requires
$f\sim2.0$ for the source population as a whole. This implies either 
a continuum spectrum which is significantly less steep than the canonical form
({ i.e.}, $\alpha\sim 0.4$) or a spectrum which is cut-off at low energies,
presumably due to X-ray absorption in relatively cold line-of-sight gas 
intrinsic to the source. 
For the canonical spectral slope, $f\sim2.0$ implies $N_H\sim 3\times 
10^{21}\rm~cm^{-2}$. This is barely consistent with the upper 
limit for photoelectric absorption found in the spectrum of the 
fluctuations (but see below). 
Nevertheless given the available information on the spectra of extragalactic 
X-ray sources (e.g. Maccacaro  et al. 1988), a solution in terms of 
absorption would seem to be the more tenable.

The EMSS results (Gioia et al. 1990b; Stocke  et al. 1991) confirm that AGN, 
namely Seyfert galaxies and QSOs, comprise the dominant source population 
at intermediate flux levels with other classes of source 
(clusters of galaxies, normal galaxies and BL Lac objects) providing 
up to a 25 per cent contribution to the 
extragalactic source counts. Since it is well established that low 
luminosity AGN preferentially exhibit intrinsic absorption (e.g. 
Reichert { et al.} 1985; Turner \& Pounds 1989),  
it would seem reasonable to infer that it is this class of source which is 
under-represented in the EMSS. The presence of such a spectral selection effect
may well account for at least part of the flattening in the luminosity 
function of AGN below $L_{X} \sim 10^{43}\rm~erg~s^{-1}$ 
reported in EMSS sample (Maccacaro { et al.} 1991). Observations 
of individual Seyfert I galaxies suggest that intrinsic absorption
is common in sources with luminosities less than about $3\times 10^{43}\ergps$
(Turner \& Pounds 1989). Also, it is well known that the more numerous
Seyfert~2 galaxies show strong absorption. Absorption by a column density of
$\sim10^{22}\psqcm$ will strongly suppress the emission in the soft band
of the EMSS (0.3 -- 3.5~keV) but have little effect on the 4 -- 12~keV band of
Ginga.

In order to test the above hypothesis we have constructed  simple 
model X-ray source counts. We calculated the cluster contribution 
taking into account the cluster
luminosity function of Edge \etal (1990), the negative evolution shown by the
clusters (Edge \etal 1990; Gioia \etal 1990a) and the observed temperature
luminosity relation (Edge \& Stewart 1991). The residual source counts
were then compared to the model predictions for AGN. We used
the AGN luminosity function of Piccinotti \etal (1982) and assumed
luminosity evolution of the form $(1+z)^C$  (see \eg Maccacaro \etal
1991; Boyle \etal 1992). An energy index of 0.7 is assumed. 
Intrinsic absorption
corresponding to a column density of $N_H=10^{22}N_{22}\psqcm$ is applied to
all the AGN with luminosities below $L_{\rm abs}$. An important parameter is
the assumed minimum luminosity of the AGN population, $L_{\rm min}$, which 
controls the total contribution of AGN to the X-ray background. In
practice we expect there to be a break to a flatter luminosity function but
here we simply truncate the luminosity function at $L_{\rm min}$. 

After searching parameter space we find that $C\sim2.5$, $N_{22}\sim1$
and $L_{\rm min}\sim 3 \times 10^{42}\ergps$,
$L_{\rm abs}\sim 10^{43}\ergps$ give total (AGN plus cluster) source
counts which agree with Piccinotti \etal (1982), 
the source counts derived from the $P(D)$ found here
and the EMSS counts (see Figure 8). 

The summed spectrum of the sources brighter than $S=10^{-12}\ergpcmsqps$ 
is shown in Fig. 9. It is seen that it is well fit over
the 3--12~keV band by a power law with $\alpha=0.76$, the softer spectra of
the clusters having steepened the total spectrum slightly. It is in 
reasonable agreement with the observed  2--10 keV spectral  fluctuations. 
Significant deviations
from a power law are only clear below 3~keV, where the absorption from the 
low-luminosity AGN becomes effective. The total AGN 
contribution at 4~keV, extrapolating the best fit model  to low fluxes and 
integrating to a maximum redshift of 2.5,
amounts to 63 per cent of the observed background intensity. The uncertainty
on this figure is difficult to quantify - it depends on $z_{\rm max}$, $C$
and other parameters not constrained by the \Gi \ fluctuation results.
While the uncertainty could, in principle, be $\sim 20$ per cent, the model
would then have the wrong spectral shape at 4 keV.  
Smaller values of $N_{22}$ do not give enough difference between the $P(D)$ and
the EMSS counts.  Larger values, or the same value extending to a higher
$L_{abs}$, start to affect the predicted fluctuation spectrum significantly. In
reality we expect that there are distributions of $N_{22}$, $L_{\rm min}$ and
$L_{\rm abs}$, so that not all low luminosity sources are affected in the same way.
Without much better samples of AGN spectra to constrain so many variables, we
cannot be more specific about the details at the present time.  We require 
simply that
the appropriately weighted means of these distributions influence the spectra and counts in the same
manner as the results quoted above.

Note that the effects of redshift and absorption mean that the 
apparent luminosity
evolution of sources in our model  appears steeper in soft X-rays than in
harder X-rays. Current estimates of the luminosity evolution of AGN based upon
soft X-ray luminosity functions are overestimated if some correction for 
luminosity-dependent, intrinsic
absorption is required and not made. This means that X-ray evolution, which without such
corrections is slower than at optical wavelengths, is even slower
than previously assumed. Whether there are similar corrections to be applied to
reddened Seyfert 2 galaxies, for example, in the optical band remains to be
seen.

\section  {CONCLUSIONS}

The main conclusions from our analysis of the spatial fluctuations 
in the XRB measured by \Gi \ are as follows:

1. The $\log N - \log S$ relation in the 2--10 keV band follows the form,
$N(S)dS=2\times 10^{-15}S^{-2.5}dS$,
down to $S\sim 5\times 10^{-13}\rm~erg~cm^{-2}~s^{-1}$.
The surface density of extragalactic X-ray sources at this flux limit is
$\sim 1$ object per square degree at which point approximately 
10 per cent of the 2--10 keV XRB intensity is directly accounted 
for in terms of discrete sources.

2. The excess variance technique provides a tight upper limit to the 
inhomogeneities in the source distribution and direct evidence 
for the large scale isotropy of the Universe out to redshifts $z\sim 1$.

3. The spectral properties of the XRB fluctuations are consistent
with the average 2--10 keV source having a spectral slope close to the 
canonical AGN value. Tentative evidence for high energy hardening is
found which might be interpreted in terms of a reflection bump present in the
sources that dominate the fluctuations. The softer spectra of clusters
contributing to the fluctuations may partly mask the absorption in the AGN
component. 
 
4. The normalisation of 2--10 keV source counts is a factor of 2 -- 3 above
that derived from the EMSS if a spectral 
conversion is assumed which ignores absorption effects. It is
proposed that the EMSS under-estimates the number of low-luminosity
AGN as a result of intrinsic absorption. The revised estimate of
the surface density of extragalactic sources is 
of particular importance when considering the confusion limit
of imaging systems operating above $\sim 2$ keV.
For an X-ray mission with with an effective resolution of 
$\Omega_{\rm eff} \sim 1 \rm~arcminute^2$, the confusion limit
will be reached at a flux of about $ 10^{-14} \rm~erg
~cm^{-2}~s^{-1}$.

{\it Postscript:} \\
The work described in this paper was completed in late 1994 but
due to a number of unfortunate compounding factors its
publication has been subject to protracted delay. In the interim
there have been major developments in the field stemming from the 
source counts measured by ROSAT and ASCA. The ROSAT source counts reach
unprecedently faint levels in the soft X-ray (0.5--2 keV) band 
revealing a high
density of sources ($>400\rm~deg^{-1}$ at $S = 2 \times10^{-15}\rm
~erg~cm^{-2}~s^{-1}$) which contribute over half of the X-ray 
background in this band (Hasinger et al. 1993; Branduardi-Raymont et al. 1994;
Georgantopoulos et al. 1996). However, in the relevant flux range, the ROSAT 
source counts are fully consistent with the earlier EMSS results;
thus our detailed comparison of the \Gi \ results with those
from EMSS is still valid. 
In the medium energy 2--10 keV X-ray band to which our fluctuation 
measurements apply, estimates have recently become available
of the discrete source detection rate in deep ASCA observations.
These preliminary results from ASCA, albeit based on limited source
statistics (e.g. Inoue et al 1996; Georgantopoulos et al. 1997), 
appear to be in very good 
agreement with the \Gi \ measurements in terms of the
normalisation of the log N - log S curve at a flux of
$5 \times 10^{-13} \rm~erg~cm^{-2}~s^{-1}$. The existence of a 
major X-ray source population characterised by either intrinsically flat 
or absorbed spectra is thus confirmed. For a review of other recent 
developments relating to our understanding of the wide-band X-ray to 
Gamma-ray extragalactic background, see Hasinger (1997).

\section*{ACKNOWLEDGEMENTS}

We thank Rees Williams for his help in modelling the {\it Ginga}  
background during the early stages of this work.
JAB acknowledges the receipt of a PPARC Research Studentship. 
ACF thanks the Royal Society for Support. Partial financial support to FJC
and XB was provided by the DGES under project PB95-0122. This work made 
use of the STARLINK node at Leicester which is funded by PPARC.

\pagebreak

\section*{REFERENCES}

\ref Boldt E., 1988. In: Physics of Neutron Stars and Black
Holes, Ed. Y.Tanaka, Univ. Acad. Press, Tokyo, 333 

\ref Barcons X., 1992, \apj, 396, 460

\ref Barcons X., Fabian A.C.,  Rees M.J., 1991,  Nature, 350,
685

\ref Boyle B.J., Griffiths R.E., Shanks T., Stewart G.C., Georgantopoulos I., 
1993, \mnr 260, 49

\ref Branduardi-Raymont G. et al., 1994, \mnr 270, 947

\ref  
Carrera F.J., Barcons X., Butcher J.A., Fabian A.C., Stewart G.C.,
Warwick R. S., Hayashida K., Kii T., 1991, 
\mnr 249, 698

\ref Carrera F.J, Barcons X., Butcher J.A., Fabian A.C., Stewart A.C.,
Toffolatti L., Warwick R.S., Hayashida K., Inoue H., Kondo, H., 1993,
\mnr 260, 736
  
\ref Condon J.J., 1974, \apj 188, 279 

\ref Edge A.C., Stewart G.C., Fabian A.C.,  Arnaud K.A., 1990, \mnr
 245, 559

\ref Edge A.C., Stewart G.C., 1991, \mnr  252, 414


\ref Fabian A.C., 1975, \mnr 172, 149

\ref Fabian A.C., Rees M.J., 1978, \mnr 185, 109 

\ref Fabian A.C., Barcons X., 1992, ARA\&A, 30, 429

\ref Georgantopoulos I., Stewart G.C., Shanks T., Boyle B.J.,
 Griffiths R.E., 1996, \mnr 280, 276

\ref Georgantopoulos I., Stewart G.C., Blair, A.J., Shanks T.,
 Griffiths R.E.,  Boyle B.J., Almaini O., Roche N., 1997, \mnr in press

\ref Gioia I.M., Henry J.P., Maccacaro T., Morris S.L., Stocke J.T.,
Wolter A., 1990a, \apj  356, L35

\ref  
Gioia I.M. Maccacaro T., Schild R.E., Wolter A., Stocke J.T., Morris
S.L.,  Henry J.P., 1990b, ApJS, 72, 567

\ref Hasinger G., Burg R., Giacconi R., Hartner G., Schmidt M.,
Trumper J., Zamorani G., 1993, A\&A, 275, 1 

\ref Hasinger, G., 1997, A\&A Suppl, 120, 607 

\ref Hayashida K. et al., 1989, PASJ,  41, 373

\ref Hayashida K., 1990, Ph.D. dissertation, Univ. Tokyo

\ref Inoue H., Kii T., Ogasaka Y., Takahashi T., Ueda Y., 1996,
In: Rontgenstrahlung from the Universe, eds. Zimmermann, H.U., Trumper J.,
Yorke H., MPE Report 263, 323

\ref Maccacaro T., Gioia I.M., Wolter A., Zamorani G., Stocke J.T., 
1988, \apj 326, 680 

\ref Maccacaro T., Della Ceca R., Gioia I.M., Morris S.L., Stocke, J.T.,  
Wolter, A., 1991, \apj 374, 117


\ref 
Marshall F.E., Boldt E.A., Holt S.S., Miller R.B., Mushotzky R.F.,
Rose L.A., Rothschild R.E, Serlemitsos P.J, 1980, \apj 235, 4

\ref Mather J.C., et al., 1990, \apj  354, L37

\ref 
Piccinotti G., Mushotzky R.F., Boldt E.A., Holt S.S., Marshall F.E.,
Serlemitsos P.J., Shafer R.A., 1982, \apj 253, 485


\ref Pounds K.A., Nandra P., Stewart G.C., George I.M.,  Fabian, A.C.,
1990,  Nature,  344, 132

\ref Pye J.P.,  Warwick R.S., 1979, \mnr 187, 905 

\ref Reichert G.A., Mushotzky R.F., Petre R., Holt S.S., 1985, \apj 296, 69 

\ref Scheuer P.A.G., 1974, \mnr 166, 329 

\ref Schwartz D.A., 1979, In: X-ray Astronomy, Eds. Baity \& Petersen, 
Pergamon, New York, 453 

\ref Schwartz D.A., Murray S.,  Gursky H., 1976, \apj 204, 315 

\ref Shafer R.A., 1983, Ph.D. dissertation, Univ. Maryland, NASA TM
85029 

\ref Stocke J.T., Morris S.L., Gioia I.M., Maccacaro T., Schild R.,
Fleming T.A., Henry J.P., 1991, \apjs 76, 813

\ref Turner, M.J.L. et al., 1989, PASJ,  41, 345

\ref Turner T.J.,  Pounds K.A., 1989, \mnr 240, 833 

\ref Warwick R.S.,  Pye J.P., 1978, \mnr 183, 169 

\ref Warwick R.S.,  Stewart G.C., 1989, In: X-Ray Astronomy, 2. 
AGN and the X-ray Background, Eds. Hunt J., Battrick B., 
ESA SP-296, Noordwijk, 727 

\ref Warwick R.S.,  Butcher J.A., 1992, In: Frontiers of X-ray Astronomy, 
Eds. Koyama K., Tanaka Y., Univ. Acad. Press, Tokyo, 641

\vfill\eject

\section{FIGURE CAPTIONS}

\medskip

{\bf Figure 1.} The distribution of background observations obtained from 
the \Gi \ database shown in galactic coordinates. The concentration in the 
bottom right hand corner corresponds to background observations made 
during the regular monitoring of SN 1987A. 


{\bf Figure 2.} The measured $P(D)$ histograms (a) 
in the 2--4~keV band and  (b)
4--12~keV band, together with the best fit model predictions (filled circles).

{\bf Figure 3.} One sigma  and 90 per cent contours in the $K,\gamma$
parameter space for (a) the 2--4~keV band and (b) the 4--12~keV band. 
$K$ is in units of  (LAC ct$\, {\rm s}^{-1})^{\gamma-1}\, \rm sr^{-1}$.

{\bf Figure 4.} Upper bounds to inhomogeneities as a function
of comoving scale based on the two-sigma upper limits to the excess 
fluctuations from {\it HEAO--1} A2 and {\it Ginga}.  It
is assumed that fluctuations come from a redshift range $\Delta z=0.5$ around
$z=0.5$.

{\bf Figure 5.} The spectrum of the fluctuations (top panel)
and the residuals (bottom panel) after fitting a model including 
a reflection component.

{\bf Figure 6.} The differential source counts in the 2--10~keV
band normalized to $N_0(S)=2\times 10^{-15} S^{-2.5}$.  Points are from
Piccinotti \etal (1982), the dashed contour is the 90 per cent
contour from HEAO--1
A2 fluctuations (Shafer 1983) and the solid curve is the 90 per cent contour
from the present fluctuation analysis.  The dotted line  
shows the EMSS prediction assuming $f=1$ (see text).
The dash-dotted line to the left represents the boundary of the region where
a particular power-law extrapolation completely produces the XRB.

{\bf Figure 7.} Curves of constant $f$ for values of photon spectral index,
$\Gamma$, and absorbing column, $N_H$, typical of AGN.

{\bf Figure 8.} The integrated source counts from the AGN model described in
the text. The upper solid line represents the fiducial model in the 2--10
keV band 
after subtracting the cluster contribution. The
lower solid line corresponds to the EMSS measurement. The dashed and
dotted lines correspond to the integrated counts from AGN  in
the hard and soft bands respectively. The flux units are 
$ \rm log(S_{2-10}) $ for the hard source counts and $ \rm log(S_{0.5-3.5}) $  for the soft counts.  

{\bf Figure 9.} The
cumulative contribution (lower solid line) of AGN (dotted line) and
clusters (dashed line) brighter than a 2--10~keV flux of
$S=10^{-12}\ergpcmsqps$ is compared with the intensity of the XRB
obtained from Gruber's (1992) formula (upper solid curve). 
The best-fitting power-law in the range 3--12~keV
to the AGN contribution is shown by the dash-dotted line. It has an
energy index of 0.76, slightly steeper than that of the underlying AGN
spectra of index 0.70.

\newpage

{\bf Table 1.}  The parameters of the best-fitting source-count models.
The normalisation values are quoted in both count rate and flux units.

\begin{table}[h]
\begin{tabular}{lcccccc} 
   & \multicolumn{3}{c}{\bf 2--4 keV Band} & \multicolumn{3}{c}{\bf 4-11 keV
Band} \\ 
   & $ \gamma$ & K~$^{a}$ & K~$^{b} $ & $ \gamma$ & K~$^{a} $ & K~$^{b}$ \\ 
Overall best-fit  & $2.53^{+0.15}_{-0.13}$ &  $210^{+53}_{-36}$ & 
$0.88^{+0.20}_{-0.15}$ & $2.58^{+0.18}_{-0.14}$ & $210^{+60}_{-38}$ 
& $0.22^{+0.07}_{-0.04}$ \\ 
Euclidean case    & 2.5  & $218^{+54}_{-36}$ & $2.0^{+0.5}_{-0.3}$ & 2.5  
& $234^{+56}_{-41}$ & $2.0^{+0.5}_{-0.3}$ \\ 
\end{tabular}
\end{table}
$^{a}$ in units of $(\rm LAC~ct~s^{-1})^{\gamma-1}~sr^{-1}$. \\
$^{b}$ in units of $10^{-15}~(\rm erg~cm^{-2}~s^{-1})^{\gamma-1}~sr^{-1}$.

\end{document}